\documentclass[aps,pra,10pt,twocolumn,amsmath,amssymb,
    longbibliography,superscriptaddress,nofootinbib]{revtex4-1}

\usepackage{graphicx}% Include figure files
\usepackage{dcolumn}% Align table columns on decimal point
\usepackage{bm}% bold math
\usepackage{braket}
\usepackage{physics}
\usepackage{amsmath}
\usepackage{amssymb}
\usepackage{makecell}

\begin{document}

%\preprint{APS/123-QED}

\title{Polarization switching induced by domain wall sliding in two-dimensional ferroelectric monochalcogenides}

\author{Urko Petralanda}
 \email{upeho@dtu.dk}
\affiliation{%
 Computational Atomic-Scale Materials Design (CAMD), Department of Physics, Technical University of Denmark, 2800 Kgs. Lyngby, Denmark
}%

\author{Thomas Olsen}%
 \email{tolsen@fysik.dtu.dk}
\affiliation{%
 Computational Atomic-Scale Materials Design (CAMD), Department of Physics, Technical University of Denmark, 2800 Kgs. Lyngby, Denmark
}%

\begin{abstract}
The ability to switch between distinct states of polarization comprises the defining property of ferroelectrics. However, the microscopic mechanism responsible for switching is not well understood and theoretical estimates based on coherent monodomain switching typically overestimates experimentally determined coercive fields by orders of magnitude. In this work we present a detailed first principles characterization of domain walls (DWs) in two-dimensional ferroelectric GeS, GeSe, SnS and SnSe. In particular, we calculate the formation energies and migration barriers for 180$^\circ$ and $90^\circ$ DWs, and then derive a general expression for the coercive field assuming that polarization switching is mediated by DW migration. We apply our approach to the materials studied and obtain
%that the switching mechanism associated with DW sliding yields coercive fields that are in 
good agreement with experimental coercive fields. The calculated coercive fields are up to two orders of magnitude smaller than those predicted from coherent monodomain switching in GeSe, SnS and SnSe. Finally, we study the optical properties of the compounds and find that the presence of 180$^\circ$ DWs leads to a significant red shift of the absorption spectrum, implying that the density of DWs may be determined by means of simple optical probes.
\end{abstract}

\maketitle

\section{Introduction} \label{section:intro}
The number of experimentally available two-dimensional (2D) materials is currently expanding at a rapid pace and there is increasing interest in functional materials based on monolayer and few-layer compounds. In particular, exfoliation of several 2D materials exhibiting robust ferroelectricity has recently been demonstrated and these materials are currently the subject of intense experimental as well as theoretical scrutiny \cite{Chang2016, Higa2020,Xiao2018,Chang2020,Chang2020Parkin,Shirodkar2014,Yuan2019,Zhou2017,Fei2018,Li2017,Liu2016,Wu2017_2,RevModPhys.93.011001}. The versatile properties of 2D materials facilitate bottom up design of van der Waals heterostructures that may be optimized with respect to desirable properties for applications such as  photovoltaics \cite{Rangel2016,Li2021} or - in the case of ferroelectrics - photoferroics \cite{Castelli2019}. It is thus pertinent to acquire a detailed understanding of individual 2D ferroelectrics and several important steps have been taken towards this over the past decade \cite{Shirodkar2014,Wang_2017,Liu2018,Ding2017,Li2017,Ruixiang2016}. However a detailed understanding of polarization switching has so far been restricted to qualitative arguments and a quantitative theory for the determination of coercive fields is completely lacking 

Polarization domains and associated DWs naturally emerge in ferroelectric crystals since they tend to minimize the depolarizing field arising from bound charges at edges and surfaces. Furthermore, when a ferroelectric is cooled below the Curie temperature different spontaneous polarization vectors may nucleate at different positions, which ultimately leads to an intrinsic domain structure. Apart from strongly influencing the electronic and dynamical properties of the bulk material, DWs constitute intriguing objects of interest by themselves. In 3D crystals they are intensively studied as building blocks for nanoelectronic devices such as memristors for neuromorphic computers \cite{Salje2021,Catalan2012,McConville2020,Matzen2014} and the recent experimental characterization and manipulation of DWs in monolayer ferroelectrics \cite{Chang2020} suggest that DWs may play a similar role in the development of devices based on 2D materials. DWs have been experimentally characterised in monolayer and few-layer ferroelectrics including SnX (X= Te, Se, S), MoTe$_2$, 
%MoX$_2$ (X=S, Te), 
CuInP$_2$S$_6$, 
%WTe$_2$ 
and In$_2$Se$_3$ \cite{Chang2016,Chang2020,Dziaugys2020,Bao2019,Yuan2019,Xu2021}. In particular, for the case of the monochalcogenides both neutral and charged (conductive) 90$^\circ$ and 180$^\circ$ DWs have been observed \cite{Chang2016, Chang2020Parkin}. Despite its paramount importance for device applications the mechanism underlying polarization switching in ferroelectric monolayers remains poorly understood and first principles calculations have reported %coherent monodomain polarization switching
coercive fields several orders of magnitude larger than the experimentally observed fields \cite{Higa2020, Hanakata2016, Wang_2017}. However, these calculations are based on coherent monodomain switching and it is well known that the process of DW destabilization (polarization switching mediated by DW migration) is likely to be much more favorable. 

In this work, we present a first principles study of neutral 90$^\circ$ and 180$^\circ$ DWs in monolayer GeS, GeSe, SnS and SnSe. We determine their structural, electronic, optical and dynamical properties and propose a parametrization of the properties in terms of the in-plane anisotropy, which allows us to extend the conclusions to other monochalcogenides. The calculations demonstrate that the coercive field associated with DW sliding is consistent with the experimentally observed coercive field in SnS and thus confirm that DW migration is indeed the dominating mechanism for polarization switching. Our DW free energy approach for the determination of coercive fields is rather universal and the formalism can thus be applied to other ferroelectrics as well.
Finally, we show that the optical absorption in the solar spectral range is strongly enhanced by DWs, which is critical in the context of photovoltaic devices.

\section{Results and Discussion} \label{section:results}

\subsection{Structural Details} \label{section:stru}

\begin{figure} 
    \includegraphics[width=0.8\linewidth]{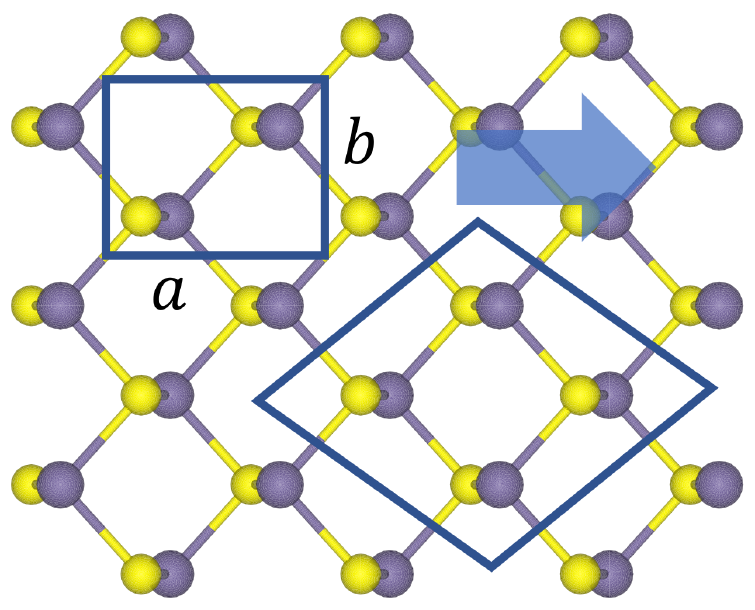}
\caption{    \label{fig:cell}
   Atomic structure of monolayer GeS. The rectangular unit cell is illustrated with lattice parameters denoted by $a$ and $b$ and the direction of spontaneous polarization is indicated by the arrow. The rhombus unit cell used for constructing 90$^\circ$ DWs is shown below.}
\end{figure}

The monolayer group IV monochalcogenides studied in this work are isostructural and are all well known experimentally as van der Waals bound 3D structures \cite{Chang2016,Higa2020,Chang2020, Chang2020Parkin, Haastrup2018,Gjerding_2021}. The minimal unit cell contains 4 atoms and has the orthorhombic space group P2$_1$am. The prototype unit cell of the compounds is shown in figure \ref{fig:cell}, where the direction of spontaneous polarization is indicated. In table \ref{tab:table3} we present an overview of various properties of the four materials, which will be described in detail below. The differences between the four materials can largely be attributed to differences in the in-plane anisotropy defined as $\delta = \frac{\abs{a-b}}{a+b}$ \cite{Vannucci2020}, which is directly correlated with the magnitude of spontaneous polarization.

The simulation of the 180$^\circ$ DW was carried out by taking a supercell of 24 minimal unit cells and displace atoms in half of the supercell such that we obtain two domains with polarization in positive and negative $y$-directions.
In order to perform a structural relaxation of the DW we fixed the lattice parameter parallel to the DW, but allowed the lattice to expand in the direction orthogonal to the DW. In the case of an isolated DW such expansion is allowed since the DW itself will introduce strain orthogonal to the DW. In contrast, there cannot be finite strain in the direction parallel to the DW since this would lead to a divergent energy contribution from the strain in the bulk. On the other hand, if one is interested in a regular array of DWs (which is what we are actually simulating) there may be a certain amount of strain in the direction parallel to the DW and we do in fact find a sizeable amount of strain if we do not fix the lattice parameter parallel to the DW (see table S1 in the Supplementary Information (SI) for more details on this issue). To determine the centering of the DW, we relaxed both a bond centered and an atom centered configuration (see figure S1 in the SI for further details). The bond centered structure has lower energy in GeS and GeSe, while the atom centered structure has a lower energy in SnS and SnSe. The energies of two latter compounds are, however, very similar. 

\begin{figure*}
    \includegraphics[width=1.0\linewidth]{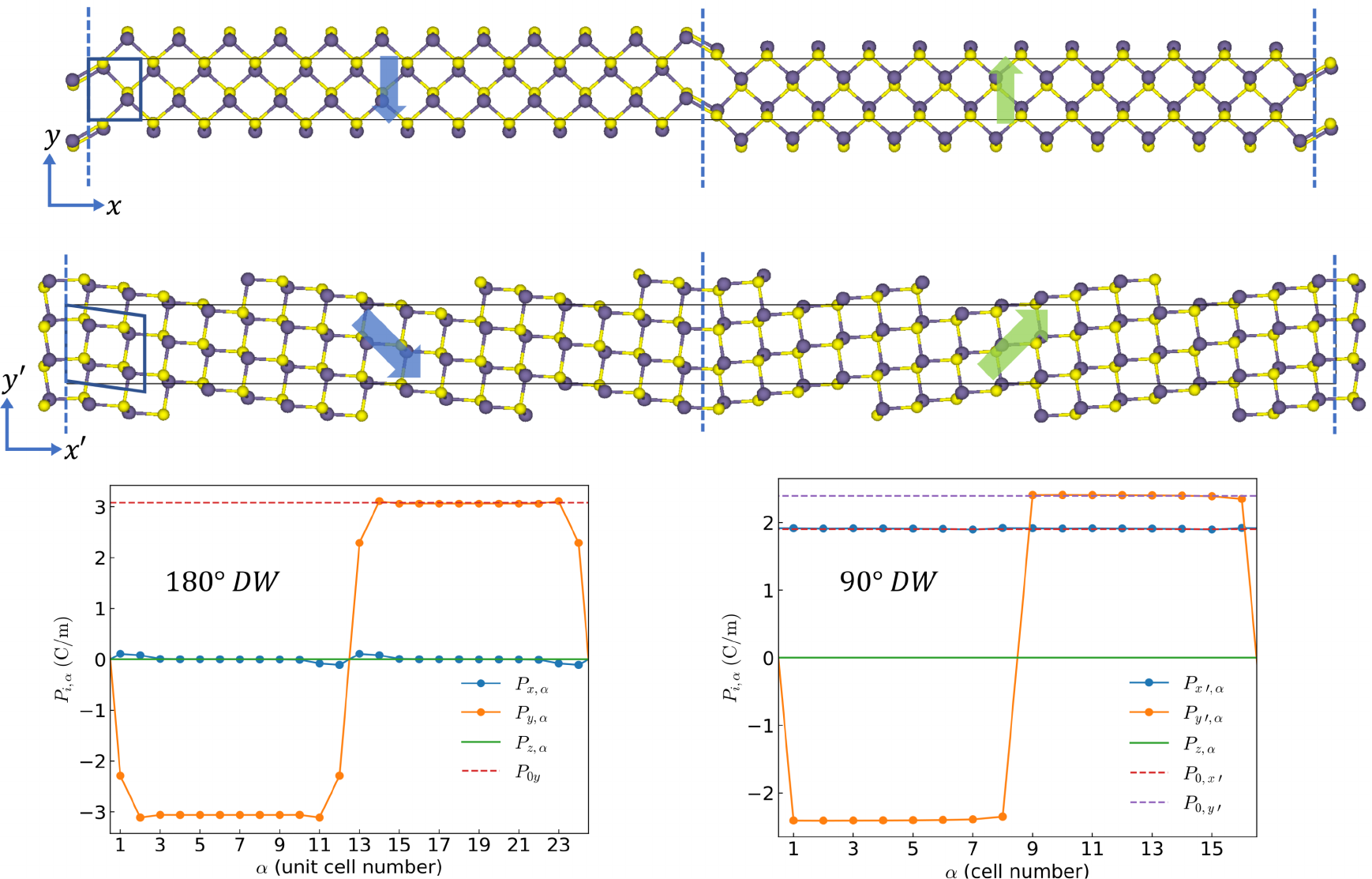}
    \caption{
        \label{fig:stru}
        The top (middle) panel shows the relaxed 
        structure of a supercell with two 180$^\circ$ (90$^\circ$) DWs in GeS. The centers of the DWs are indicated by dashed lines and the directions of polarization in the domains are indicated by arrows. The bottom panel shows the spontaneous polarization profiles of the corresponding supercells. The polarization is shown along the axis of the coordinate systems shown next to the two supercells. The components of the \emph{pristine} monolayer polarization $P^0_i$ in each of the two systems are shown for reference. The minimal and rhombus cells defined in figure \ref{fig:cell} are also shown for clarity.}
\end{figure*}

In the case of 90$^\circ$ DWs, we first constructed a rhombus unit cell with polarization along its long diagonal (see figure \ref{fig:cell}). We then built a supercell consisting of 16 rhombi cells, which are situated such that we obtain two domains with polarization vectors rotated by an angle close to 90 degrees with respect to each other (see  figure \ref{fig:stru}). The polarization vectors in the two domains are aligned head to tail such that we maintain DW bound charge neutrality. 
The anisotropy of the minimal unit cell yields an angle $\gamma$ between polarization vectors of the two domains, which is slightly smaller than 90 degrees and is given by $\tan(\gamma/2)=b/a$. 
We then conducted a full relaxation of these supercells in all compounds where we again fixed the lattice parameter parallel to the DW. In contrast to the case of 180$^\circ$ DWs, however, the lattice stress emerging after relaxation is very small, indicating that 90$^\circ$ DWs perturb the compounds significantly less compared to 180$^\circ$ DWs. We finally compared the energies of bond and atom centered DWs and found that the former resulted in the most stable 90$^\circ$ DWs in all four materials.

In order to calculate the polarization profile in the DW supercells  we adapted the approach from ref \cite{Meyer2002} and write the polarization in the $i$'th direction of unit cell $\alpha$ as
\begin{align}
P_{i,\alpha}=\frac{1}{A}\sum_{j,a\in\alpha}Z_{ij}^{*a} d_j^a, 
\end{align}
where $A$ is the unit cell area, $d^a_j$ is the displacement of atom $a$ in direction $j$ and $Z_{ij}^{*a}$ is the Born effective charge tensor of atom $a$. We note that the Born effective charges may exhibit sizeable variations between the centrosymmetric and polar structures and we have used the average of these two values in order to obtain a compromise that provides a reasonable description of the polarization in the vicinity of the DW (see table S2 in the SI for a comparison of the Born effective charges in the centrosymmetric and polar phases). In figure \ref{fig:stru} we show the polarization profiles for 90$^\circ$ and 180$^\circ$ DWs of GeS. The polarization reaches the bulk value one and two unit cells from the DW for 90$^\circ$ and 180$^\circ$ DWs respectively and the DW width is thus at most a few unit cells wide in both case in both cases.

\begin{table}[t]
\resizebox{0.47\textwidth}{!} {%
\begin{tabular}{lllll}
\hline
formula &   SnSe  &    SnS  &   GeSe  &   GeS  \\
\hline
$\delta$ (calculated)     &   0.014 &    0.031 &    0.036 &    0.10 \\
$\delta$(exp.) \cite{Chang2020}&   0.011  &    -  &   -  &    -  \\
$\delta$(bulk exp.) \cite{Larsen2019,Chattopadhyay1986NeutronDS,Wyckoff1963,Bissert1978}&   0.029 &    0.014 &   0.066 &    0.083  \\
$w^{}_{180^\circ}$[\r{A}]     & 22.4  & 11.7   & 10.5  & 6.05    \\
$w_{90^\circ}$[\r{A}]     &    8.11 &    5.99 &    6.19 &    5.95 \\
%$\alpha_{90^\circ}\;[^{\circ})$    &    1.66 &    3.57 &    4.17 &    11.9 \\
$\gamma\;[^{\circ}]$     &      90.0 &    86.5 &   86.7 &     78.2 \\
%centering    &  0.0343 &   0.187 &  0.188 &  0.0681 \\
\hline
$\Delta \varepsilon_{180^\circ}$[eV]     &  -0.11 & -0.14 & -0.24  & -0.34  \\
$\Delta \varepsilon_{90^\circ}$[eV]     & 0.01  &  0.03 & -0.02  & 0.03  \\
$a_{180^\circ}[\%]$     &    1.9 &    12 &    2.5 &   40 \\
$a_{90^\circ}[\%]$    &    0.33 &  -0.1 &   -0.05 &     7.4 \\
\hline
$E_{180^\circ}^\mathrm{F}$[meV/\r{A}]     &   22.8 &    53.6 &     75.5 &     145 \\
$E_{90^\circ}^\mathrm{F}$[meV/\r{A}]     &   8.13 &   19.2 &   16.4 & 25.0  \\
$E^\mathrm{B}_{180^\circ}$[meV/\r{A}]    &  0.02 & 0.01 &  0.06 &  24.25 \\
$E^\mathrm{B}_{90^\circ}$[meV/\r{A}]     &  0.3 &  4.9 &  4.5 &  27.3 \\
$\mathcal{E}_\mathrm{C,180^\circ}$[kV/cm]    &  11   &   2.6 &   22   &   7.5$\times10^4$ \\
$\mathcal{E}_\mathrm{C,90^\circ}$[kV/cm]    &  170   &   2.2$\times10^3$ &  1.6$\times10^3$   &   7.0$\times10^3$ \\

\end{tabular}}
\caption{\label{tab:table3} Properties of SnSe, SnS, GeSe and GeS monolayers. Values of the lattice anisotropies $\delta$, DW width $w$, polarization angle at the 90$^\circ$ DWs $\gamma$, band gap change $\Delta \varepsilon_\mathrm{gap}$, absorption gain under solar illumination $a$, DW formation energy per unit length $E^\mathrm{F}$, DW migration energy barrier per unit length $E^\mathrm{B}$ and coercive field for DW migration $\mathcal{E}_\mathrm{C}$ calculated for each DW type.}
\end{table}

\subsection{Stability} \label{section:stability}
The formation energy per unit length of a DW is given by
\begin{align}
E^\mathrm{F}=\frac{1}{2a_\parallel}(nE_{0}- E_{\mathrm{ DW }}),    
\end{align}
where $n$ is the ratio between the number of atoms in the supercell and the number of atoms in the minimal unit cell, $a_\parallel$ is the supercell lattice parameter parallel to the DW, $E_0$ is the calculated monolayer energy per unit cell and $E_\mathrm{DW}$ is the calculated energy of the supercell with two DWs. The values obtained for all compounds are shown in table \ref{tab:table3}. The 90$^\circ$ DWs are more stable in all cases 
and we observe that a larger polarization and thus a larger anisotropy of the pristine monolayer implies a smaller DW width and a larger formation energy of the two types of DWs (see table \ref{tab:table3}). This trend is in good agreement with the values in Ref. \cite{Wang_2017}, although our formation energies are higher, especially in 180$^\circ$ DWs. The reason for this is the strong strain parallel to the DWs that emerges under a full relaxation of the supercell as mentioned above.  In the SI we show the formation energies of 180$^\circ$ DWs obtained for a full relaxation, which agree well with Ref. \cite{Wang_2017}.

\subsection{Dynamical Properties} 

The possibility of switching polarization by means of an external electric field comprises the defining property of ferroelectric materials. There are multiple factors influencing the polarization switching, including point and structural defects, strain fields or device specific factors \cite{Scott2016}. The importance of these effects are difficult to distinguish experimentally and first principles simulations thus become vital tools for gaining insight into the microscopic mechanisms underling the switching process. 
Since ferroelectric crystals in general are not monodomain systems, the domain structure and size of the crystal will determine the optimal switching path of the material. In practice, once the crystal is large enough to host DWs, the polarization switching has been shown to occur locally through DW migration/nucleation in 3D crystals \cite{Merz1954,Jung2002,Huang2016,Rubio2015}. In 2D ferroelectrics this has not yet been experimentally verified although it seems likely that DWs could play a prominent role in polarization switching. The previously calculated coercive fields for monochalcogenides are limited to coherent monodomain switching \cite{Hanakata2016, Wang_2017} and these values exceeds the experimental data in 2D SnS by at least two orders of magnitude \cite{Higa2020}. Here we calculate the coercive field associated with polarization switching originating from DW migration and find that it is in much better agreement with the experimentally observed fields.

\subsubsection{Migration Barriers}

\begin{figure*}
    \includegraphics[width=1.0\linewidth]{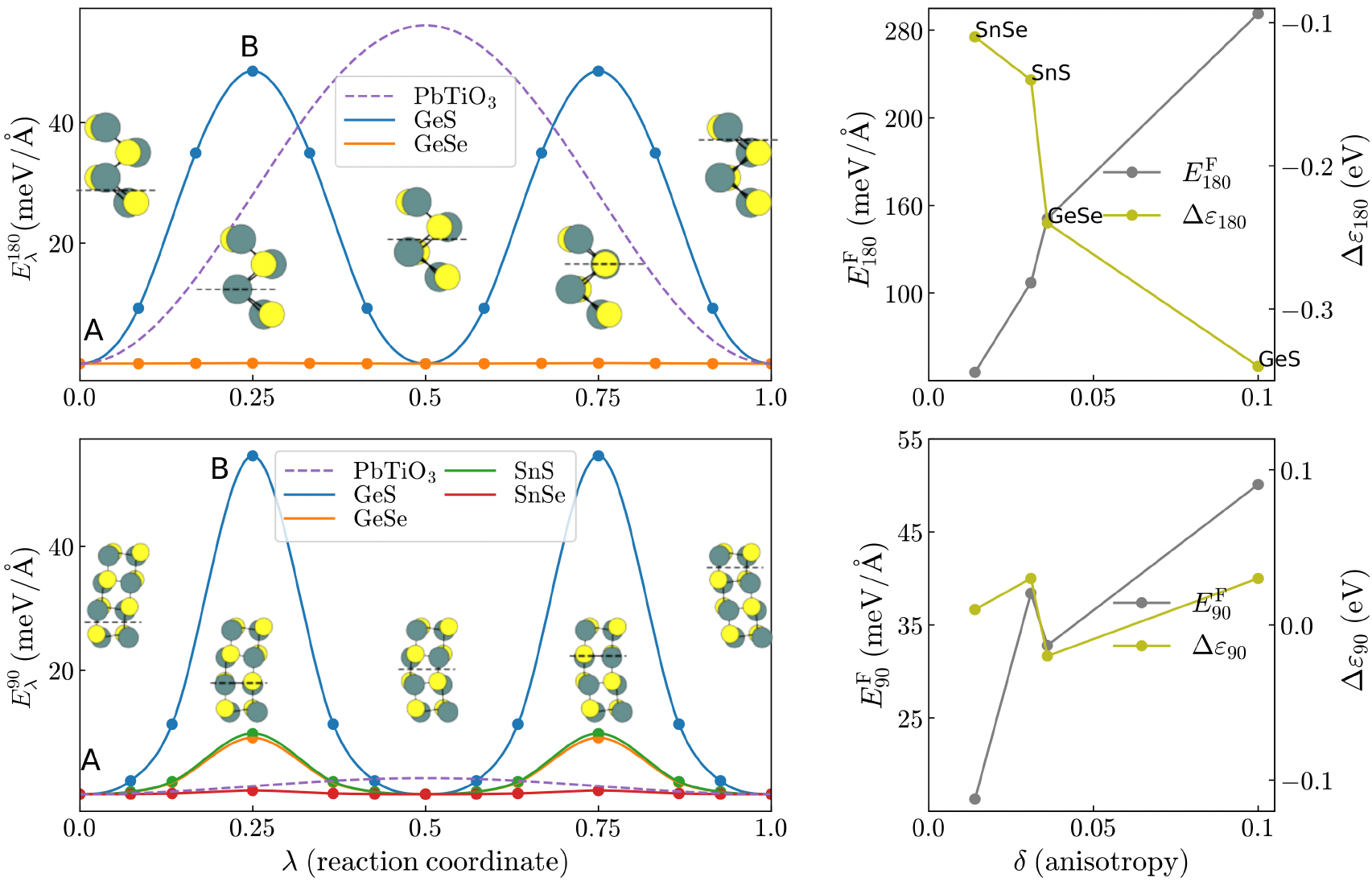}
    \centering
    \caption{
        \label{fig:switch}
        Top (bottom) left: the DW migration energy profiles for 180$^\circ$ (90$^\circ$) DWs. For the 180$^\circ$ case the curves for SnS and SnSe are not shown since their energy barriers are negligible. For comparison, we also show the curve for a 1 nm slab of PbTiO$_3$ (adapted from Ref. \cite{Meyer2002}). Top (bottom) right: formation energies and band gap changes of 180$^\circ$ (90$^\circ$) DWs for the four compounds ordered by lattice anisotropy.}
\end{figure*}

DW migration energy profiles can be calculated using the Nudged Elastic Band (NEB) method \cite{Mills1994,Johnsson1998}, which has previously been applied to other ferroelectrics to determine coherent \cite{Beckman2009,Dogan2019} or DW mediated \cite{Li2018} energy barriers for switching. We consider the initial configurations shown in figure \ref{fig:stru} and final configurations where the DW has been shifted by one unit cell. Since the supercells are rather large a full relaxation is computationally demanding, but there are certain symmetry considerations that can be exploited to significantly decrease the cost of the NEB calculations. We start by running preliminary NEB calculations using small supercells in GeS; one with two 180$^\circ$ DWs and 32 atoms and another one with 90$^\circ$ DWs and 48 atoms. In both cases we observe that the structure halfway along the NEB path has the same symmetry and energy as the initial and final images. This was already noted by Wang \emph{et al} in Ref. \cite{Wang_2017} for 180$^\circ$ DWs in GeS. In addition, the energy along the reaction path between the initial (denoted by A) and the intermediate configuration described above (DW shifted by half a unit cell) is even around the energy barrier. Denoting the maximum energy structure by B, we can then perform a relaxation of the structure B (constrained by symmetry) and restrict the NEB calculation to points between A and B. In figure \ref{fig:switch} we sketch the optimized structures along the path. The 180$^\circ$ DWs and 90$^\circ$ DW barriers are presented in Table \ref{tab:table3} for the four materials.

The energies along the optimized NEB paths for the two DW types are shown in figure \ref{fig:switch} for all the compounds and we include the equivalent energy path for PbTiO$_3$ (assuming a layer thickness of 1 nm) \cite{Meyer2002}) for comparison. A dramatic increase of the reaction barriers with increasing anisotropy is observed for both types of DWs. In the case of 180$^\circ$ DWs, the reaction barriers for GeSe, SnS and SnSe are exceedingly small (less than 0.1 meV/\AA), while the reaction barrier of GeS is almost two orders of magnitude larger (48.5 meV/\AA). The reaction barrier of GeS is thus even larger than a 1 nm slab of PbTiO$_3$. For the 90$^\circ$ DWs we also observe an increase in barriers with anisotropy although the differences are less dramatic and varies from 0.3 meV/\AA (SnSe) to 27.3 meV/\AA (GeS). The experimentally investigated monolayer ferroelectrics (SnS and SnSe), are thus predicted to have very low reaction barriers for DW sliding. In Ref. \cite{Wang_2017} the DW migration barrier of a 180$^\circ$ DW in GeS was reported to have a barrier of 1.6 meV/\AA, which is much lower than we find. The reason for the disagreement is due to the strong strain parallel to the DWs that appears under a full relaxation of the supercell, which we avoid by fixing the lattice parameter parallel to the DW. In the SI we show the energy barrier obtained for a full relaxation in GeS, which is an order of magnitude smaller and agrees well with Ref. \cite{Wang_2017}. We stress that the supercells used for simulating DWs in real materials represent a rather high DW density and in order to obtain a faithful representation of an isolated DW is is important not to allow for strain in the direction parallel to the DW. In the case of 90$^\circ$ DWs, the DW introduces a much weaker perturbation of the system and does not introduce any sizable strain effects. 

While the 180$^\circ$ and 90$^\circ$ DW migration paths are qualitatively similar, the 90$^\circ$ NEB profiles are more sharply peaked. In the low anisotropy systems, which include the experimentally investigated monolayers SnSe and SnS, we thus observe a rather small curvature in the energy at the minimum. This implies that it should be possible to excite a DW sliding mode of very low frequency, typically in the gigahertz range, by an externally applied alternating current (AC) field. A similar mode has been reported for YMnO$_3$ \cite{Wu2017}.

We finally note that there are fundamental differences between the mechanism involved in 180$^\circ$ and 90$^\circ$ DW sliding. Specifically, sliding the 180$^\circ$ DW essentially involves only local atomic displacements in the vicinity of the DW. In contrast, the anisotropy and resulting non-equivalent angles of the rhombus cell implies that the sliding of 90$^\circ$ DWs can only happen through a concurrent shift of an entire adjacent domain. In general, DWs are macroscopic objects that cannot move coherently by thermal processes. Instead thermal fluctuations are expected to introduce DW wiggles (and associated strain) that may eventually lead to thermally activated stepwise DW migration. However, for the case of 90$^\circ$ DWs, there will be an additional large entropic barrier associated with a coherent shift of the adjacent domain and this may provide a topological blockade for thermal 90$^\circ$ DW migration.

\subsubsection{Coercive Fields}

The coercive field $\bm{\mathcal{E}}_\mathrm{C}$ is the electric field required to switch the state of polarization in a ferroelectric. We may consider two limiting mechanisms under which the polarization can be switched. One is the uniform switching of the polarization, which is largely determined by the maximal slope of the barrier that connects two opposite orientations of the spontaneous polarization. The second mechanism, which will be treated here, is based on the migration of a DW through the compound and the coercive field is then determined by the barrier for shifting a DW by one unit cell. We consider an extended system with a single DW in the presence of a constant electric field $\bm{\mathcal{E}}$. In general we may write the electric enthalpy \cite{Souza2002} per unit cell parallel to the DW as
\begin{equation}
F(\{u_i^a\},\bm{\mathcal{E}}) = E(\{u_i^a\}) -\frac{1}{a_\parallel}\sum_i\mathcal{E}_iP_i(\{u_i^a\}),
\label{eq:total_energy}
\end{equation}
where $P_i=e\sum_{a,j}Z_{ij}^*u_j^a$ is the total dipole moment of the system and $a_\parallel$ is the unit cell length in the direction parallel to the DW. $u_i^a$ denotes the position of atom $a$, $e$ is the electronic charge and $Z_{ij}^a$ is the Born effective charge tensor of atom $a$. $E[\{u_i^a\}]$ is the electronic ground state energy per unit length with the atoms fixed at positions $u_i^a$.

As the DW moves along the minimum energy path (as determined by a NEB calculation), we can parameterize the displacement by a dimensionless parameter $\lambda$ proportional to the total dipole moment along the path such that

$\lambda=0$ corresponds to the initial state and $\lambda=1$ to the final state. Each value of $\lambda$ is characterized by a set of atomic positions $u_i^{a,\lambda}$ and thus a dipole moment $P_i(\lambda)$. We can then write the electric enthalpy along the path as
\begin{equation}
F[P^\lambda,{\mathcal{E}}] = E(\lambda)-\frac{1}{a_\parallel}\mathcal{E}P_\mathcal{E}(\lambda), 
\label{eq:enthalpy2}
\end{equation}
where $P_\mathcal{E}(\lambda)$ is the component of the total electric dipole moment of the system parallel to the external electric field. 

The equilibrium configuration for a given electric field is determined by imposing that the derivative of $F$ with respect to $P^\lambda$ vanishes (see more details in the SI).
The coercive field corresponds to the smallest electric field required to overcome the energy barrier. This will happen when the force exerted by the electric field matches the maximum force along the reaction path % with respect to the total polarization 
\cite{Kim2002}. It may be expressed as
\begin{equation}
\mathcal{E}_\mathrm{C}=a_\parallel\frac{\mathrm{max}\bigg[ E'[\lambda]\bigg] }{\Delta P_\mathcal{E}},
\end{equation}
where $E[\lambda]$ is the NEB energy profile, shown in figure \ref{fig:switch}, $\Delta P_\mathcal{E}$
is the change in dipole moment along the field between the initial and final states and the prime indicates the derivative with respect to $\lambda$. (See SI for a detailed derivation.) 
$\Delta P_\mathcal{E}$ depends on the orientation of the electric field with respect to the DW and the minimal coercive field is obtained when the field is oriented parallel to the spontaneous polarization. For example, in the case of 180$^\circ$ DWs we have $\Delta P^{180^\circ}_\mathcal{E} = 2ab \cos{\theta}P_0^{2D}$, where $P_0^{2D}$ is the spontaneous 2D polarization and $\theta$ the angle between the electric field and the DW. 

In practice, we calculate the polarization resolved in unit cells using the Born effective charges, as described in Section \ref{section:stru}. We approximate the Born effective charges as being constant along the reaction path. While the Born effective charges do in fact change along the path, the polarization agrees reasonably well with exact calculations (see table S1 in the SI for the Born effective charges and polarization calculated with this method for all compounds). We have evaluated the expression above directly from the NEB energy profiles and the spontaneous polarization for the four materials considering both 90$^\circ$ and 180$^\circ$ DWs. 
The values are stated in table \ref{tab:table3}. In the case of SnS we find $\mathcal{E}_\mathrm{C}=2.6$ kv/cm, which may be compared to the experimental value of 24 kV/cm \cite{Higa2020}, obtained for a 9 layer slab. To our knowledge, the only first principles estimate for the coercive field is based on coherent monodomain switching and is reported to be 1.8$\times10^{3}$ kV/cm \cite{Hanakata2016}. The large discrepancy between this value and the experimental value was attributed to a) the fact that polarization switching does not actually happen simultaneously in all unit cells but through DW sliding or b) a potential effect of lattice strain due to the substrate \cite{Higa2020}. The present calculations suggest that the experimentally observed coercive field for SnS is fully consistent with theory if the polarization switching is assumed to be governed by DW sliding.  

Although our calculated coercive field is almost an order of magnitude smaller than the experimental value, we note that the calculated barriers and coercive fields are highly sensitive to the anisotropy, which is severely underestimated by the present calculations for SnS (see table \ref{tab:table3}). If we regard the anisotropy to be the determining parameter for the barrier, we may consider the case of SnSe instead, which our calculations predict to have an anisotropy close to the experimental value of SnS. In that case we obtain 11 kV/cm, which is much closer to the reported experimental value for the coercive field. It is, however, also possible that the presence of defects tend to pin the DWs and thus increase the measured coercive fields. In any case, our results support the view that polarization switching occurs through DW sliding and not through coherent monodomain switching in monolayer GeSe, SnS and SnSe. Further measurements and calculations should help elucidate the effect of DW pinning by defects. Interestingly, in GeS which is the most anisotropic material, the calculated coercive field due to DW sliding is similar to the coercive field for coherent monodomain switching ($6\times10^3$ kV/cm) \cite{Wang_2017} and the switching mechanism in that material could be different from the other three compounds studied here. One should, however, be cautious since the anisotropy is overestimated by the functional applied in the present work (see section \ref{section:dft}) and the coercive field due to DW sliding could thus be severely overestimated as well.

\subsection{Electronic and Optical Properties} \label{sec:elec}
\begin{figure*}
    \includegraphics[width=0.8\linewidth]{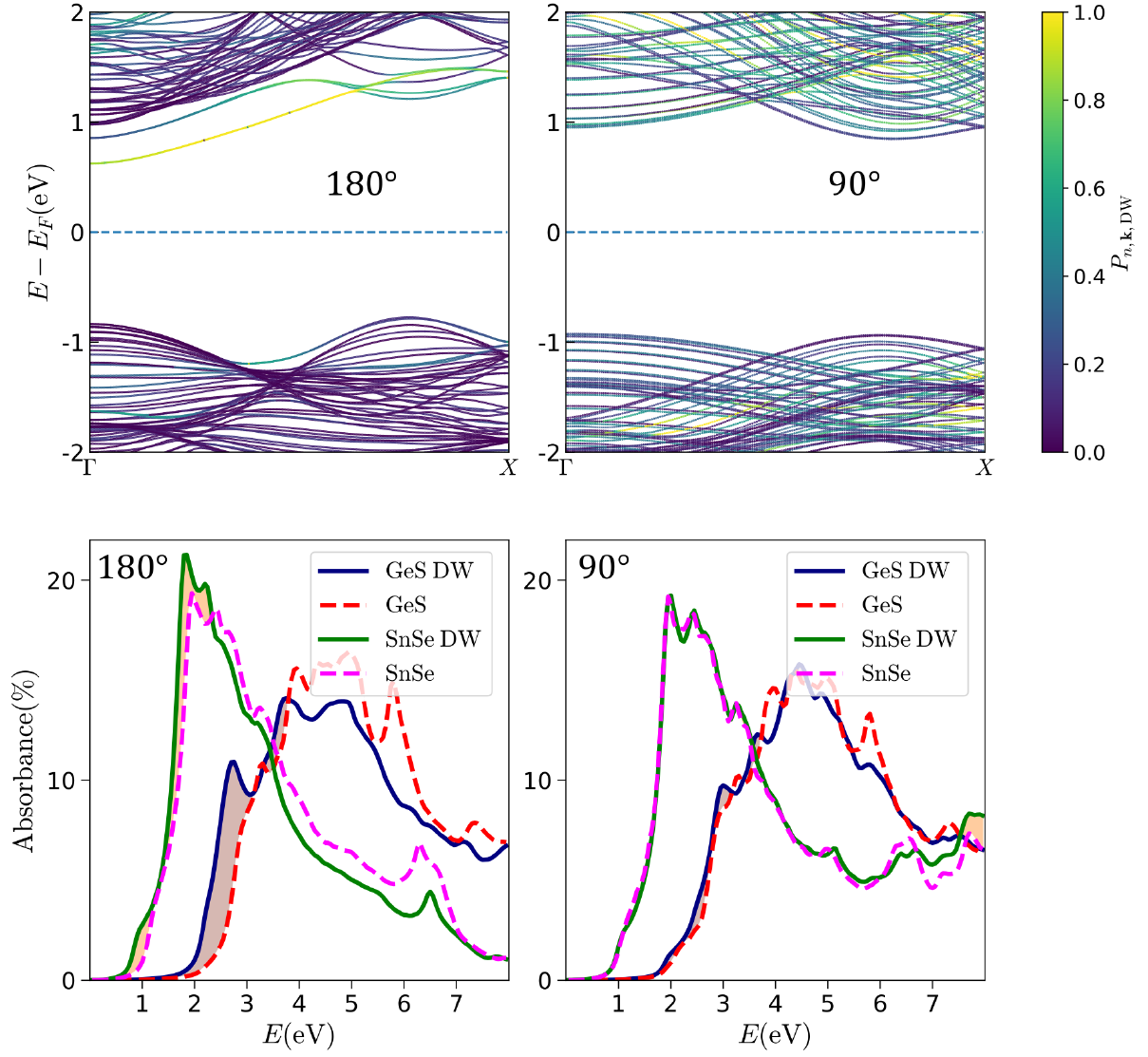}
    \centering
    \caption{
        \label{fig:electron}
        Top left (right): band structure of the relaxed 24 (16) GeS minimal (rhombus) unit cell containing two neutral 180$^\circ$ (90$^\circ$) DWs. The degree to which the states are localized in the DWs is represented according to the colorbar at the top right. Bottom: absorption spectra of Ges and SnSe, with light polarized perpendicular to the DW compared with absorption of the pristine systems. Left: $8\times1\times1$ 180$^\circ$ DW systems. Right: $6\times1\times1$ 90$^\circ$ DW systems.}
\end{figure*}

The presence of DWs is expected to influence the electronic properties and it has been demonstrated that DWs may drastically enhance the performance of photovoltaic devices based on ferroelectrics \cite{Yang2010}. At the top of figure \ref{fig:electron} we show the band structure of GeS, colored according to the degree of localization at the DWs. The localization is calculated as $P_{n\mathbf{k}}=\sum_{i, a \in DW} |\langle\phi^a_i|\psi_{n\mathbf{k}}\rangle|^2$, where $\psi_{n\mathbf{k}}$ is the Kohn Sham orbital at band $n$ and $k$-point $\mathbf{k}$ and $\phi^a_i$ is atomic orbital $i$ of atom $a$. The 180$^\circ$ (90$^\circ$) DW is taken as 4 (8) atoms wide. For the 180$^\circ$ DW we observe the appearance of a single (doubly degenerate) conduction band that is split from the conduction band manifold and completely localized at the DW. This band lowers the band gap by 0.34 eV compared to the pristine system. The 90$^\circ$ DW has a much weaker influence on the electronic structure and only gives rise to a marginal change of the band gap. In table \ref{tab:table3} and figure \ref{fig:switch} we present the change in band gap for GeS GeSe, SnS and SnSe and it is clear that the 90$^\circ$ DWs have a much smaller influence on the band structure compared to the 180$^\circ$ DWs. In addition, we observe that the change in band gap is strongly correlated with the anisotropy $\delta$.

2D materials have attracted a significant amount of attention due to their strong absorption and power conversion efficiencies compared to existing ultrathin solar cells \cite{Bernardi2013}. In addition, a sizable bulk photovoltaic effect has been predicted for ferroelectric 2D materials \cite{Yang2010,Rangel2016,Wang2017NL}. The direct calculation of the shift current in the presence of DWs is beyond the scope of this work, but the magnitude is expected to be correlated with linear optical absorption, which we investigate instead. We have thus calculated the absorption spectrum for the four pristine monochalcogenides as well as in the presence of 90$^\circ$ and 180$^\circ$ DWs. The fraction of absorbed light incident perpendicular to the atomic plane and polarized in direction $i$ can be calculated from \cite{Li_2018, Gjerding_2021}
\begin{equation}
A_i(\omega)=\Re\{\sigma^{2D}(\omega)\eta_0\}\biggr\rvert\frac{2}{2+\sigma^{2D}_i(\omega)\eta_0}\biggr\rvert^2,
\end{equation}
where $\sigma^{2D}_i(\omega)$ is the $i$'th diagonal component of the 2D optical conductivity tensor and $\eta_0$ is the vacuum impedance. The optical conductivity is calculated in the random phase approximation (RPA).

In the bottom left (right) of figure \ref{fig:electron} we show the absorption spectra of light polarized perpendicular to the 180$^\circ$ (90$^\circ$) DW in GeS and SnSe and compare with the absorption of the pristine materials (the spectra for GeSe and SnS can be found in figure S2 of SI). Note that the 180$^\circ$ and 90$^\circ$ DW planes are not parallel to each other and the pristine monolayer absorption reference is thus different in the two cases. The DW configurations correspond to high DW densities with one DW per 8 unit cells (0.56 nm$^{-1}$ DW density in GeS) for the 180$^\circ$ case and one per six rhombus cells (0.52 nm$^{-1}$ DW density in GeS) for the 90$^\circ$ case. We see that DWs give rise to a significant lowering of the absorption edge for 180$^\circ$ DWs, which is due to the decreased band gap originating from conduction band states localized at the DWs. A similar increase in absorption is observed for the 90$^\circ$ DWs, but much less pronounced. The spectra are reduced at higher frequencies as enforced by the $f$-sum rule, but for the case of solar light absorption the power conversion efficiency largely favors absorption at lower frequencies. In particular, we may integrate the absorbance weighted by the AM1.5G solar spectral irradiance ($J_{\mathrm{sol}}(\omega)$) to calculate the photon flux in the four cases according to $a_{DW}=e\int^{\infty}_{E_g}A(\omega)J_{sol}(\omega)d\omega$. By calculating the ratios of absorption in the presence DWs with respect to the pristine system, we can obtain the absorption gain for each DW type and material. We find a very high absorption gain of 40 \% for GeS 180$^\circ$ DWs for light polarized perpendicular to the DW, while for the case of SnSe the gain is merely 2\%. The 90$^\circ$ DWs gives rise to an increased absorption in the visible range as well, but the gain is much more modest compared to the 180$^\circ$ case - especially for GeS. All values for the relative gain can be found in table \ref{tab:table3}.

The calculations indicate that neutral DWs are beneficial for absorption, but they could be highly detrimental to the carrier mobilities in the direction perpendicular to the DWs. For carrier transport parallel to DWs it is, however, not obvious that mobilities are inhibited and the present calculations suggest that the photovoltaic performance of ferroelectrics could be strongly enhanced by introducing highly ordered DW structures parallel to the transport direction. In any case, our calculations imply that absorption measurements could be used as a simple probe for estimating the DW density within a given material.

Finally we note that the present calculations do not include excitonic effects, which is known to be extremely important in 2D materials. Inclusion of excitonic effects requires a Bethe-Salpether equation approach \cite{Huser2013b,Chan2021}, which is not computationally feasible for large supercells. Nevertheless, the change in absorption \textit{relative} to the pristine system is expected to be well represented by the RPA calculations performed here.

\section{Summary and Outlook}
We have studied the structural, dynamical and optical properties of neutral 90$^\circ$ and 180$^\circ$ DWs in GeS, GeSe, SnS and SnSe by means of first principles calculations. In particular, we have developed a method to calculate the coercive field associated with DW migration and find that the calculated fields are orders of magnitude smaller compared to what is obtained from coherent monodomain switching. The calculated values are in much better agreement with experimental values and it seems highly likely that DW sliding is by far the dominating mechanism responsible for switching the polarization in response to an external electric field.

Comparing the four materials, we find significant differences with respect to stability and sliding barriers of DWs. The differences are strongly correlated with the lattice anisotropy of the materials, which seems to comprise a unifying descriptor that governs stability and barriers. Large values of anisotropy lead to large formation energies and barriers for DW sliding. For all the materials, the formation energy of 90$^\circ$ DWs is smaller than that of 180$^\circ$ DWs. In general the 90$^\circ$ DWs comprise smaller perturbations of the materials, which is related to the fact that the polarization changes across DWs are smaller compared to the 90$^\circ$ DW case. This is also reflected by the fact that changes in band structures are much smaller for the 90$^\circ$ DWs than for the case of 180$^\circ$ DWs. In contrast, the DW sliding barriers are largest for the 90$^\circ$ DWs and the 180$^\circ$ DWs leads to significantly lower coercive fields (except for GeS).

Finally, we have investigated the influence of DWs with respect to optical absorption and found that the band gap reduction induced by 180$^\circ$ DWs leads to an increase of optical absorption in the solar spectral range. In particular, for the case of GeS high DW densities yield an increase in absorption of nearly 40\%. This effect is, however, more modest in less anisotropic monolayers like the experimentally scrutinized cases of SnS and SnSe. While it may be challenging to synthesize photovoltaic devices based on ferroelectrics with DWs packed at will, our findings constitutes an intriguing avenue for further theoretical and experimental exploration.

\section{Methods} \label{section:compu}
\subsection{Computational details} \label{section:dft}
Calculations were performed in the framework of density functional theory as implemented in the GPAW electronic structure package \cite{Enkovaara2010,HjorthLarsen2017}, which is base on the projector-augmented wave method \cite{Blochl1994}. We have used the Perdew-Burke-Ernzerhof exchange correlation functional \cite{Perdew1996} and a plane wave basis with an energy cutoff of 650 eV. For total energy calculations we used Monkhorst-Pack $k$-point grids with one $k$-point in the direction perpendicular to the DW, except in the smaller supercells of 32 and 48 atoms, where we used 2. In the direction parallel to the DW we sampled 25 $k$-points in the 180$^\circ$ DW supercells and 15 $k$-points in the 90$^\circ$ DW supercells. The atomic structures were relaxed until all forces were below 0.5 meV/\AA and the self-consistency was assumed fulfilled when total energies changed less than 10$^{-7}$ eV per electron in subsequent iterations. In the case of SnS and SnSe, a tighter relaxation threshold was demanded due to the small energy difference between structures and we used 0.1 meV/\AA\;for atomic relaxations and 10$^{-9}$ eV per electron for the self-consistency. For the optical absorption calculations we applied the Random Phase Approximation (RPA) and a plane wave cutoff of 50 eV for the density response function. In order to obtain smooth absorption curves we used significantly higher $k$-point densities of $60\times 60\times 1$ - $75\times 75\times 1$ for the pristine systems, and a corresponding $k$-point density in the larger DW supercells.

Applying the symmetry considerations described above, we used 4 NEB images between the A and B points and fitted the profile to a fourth order polynomial. 

\section{Acknowledgement}

The authors were supported by the Villum foundation, Grant No. 00028145.

\section*{References}

\providecommand{\noopsort}[1]{}\providecommand{\singleletter}[1]{#1}%

\clearpage
\newpage

\widetext
\begin{center}
\textbf{\large Supplementary Information for: Polarization switching induced by domain wall sliding in two-dimensional ferroelectric monochalcogenides}
\end{center}

\setcounter{equation}{0}
\setcounter{section}{0}
\setcounter{figure}{0}
\setcounter{table}{0}
\setcounter{page}{1}
\makeatletter
\renewcommand{\theequation}{S\arabic{equation}}
\renewcommand{\thefigure}{S\arabic{figure}}
\renewcommand{\thesection}{S\arabic{section}}
\renewcommand{\thetable}{S\arabic{table}}
\renewcommand{\bibnumfmt}[1]{[S#1]}
\renewcommand{\citenumfont}[1]{S#1}

\section{Structural details}

The spontaneous polarization shows a high correlation with the lattice anisotropy (see table \ref{tab:tableSI}). The Born effective charges (BEC) change substantially across the path that connects the centrosymmetric and polar phases. We therefore use the average value between these two points to calculate the polarization profiles, as indicated in the main text. The BEC values in the relevant unit cell directions are given in table \ref{tab:tableSI}, along with the spontaneous polarizations calculated using the average BEC (see main text for the expression of polarization in terms of BEC).

In order to ensure that our supercells contain a large enough DW-DW distance we conducted a convergence test at different supercell sizes. In table \ref{tab:table4} we show the convergence of energy and structural parameters of GeS as the supercell size is increased.

\begin{figure*}
    \includegraphics[width=0.99\linewidth]{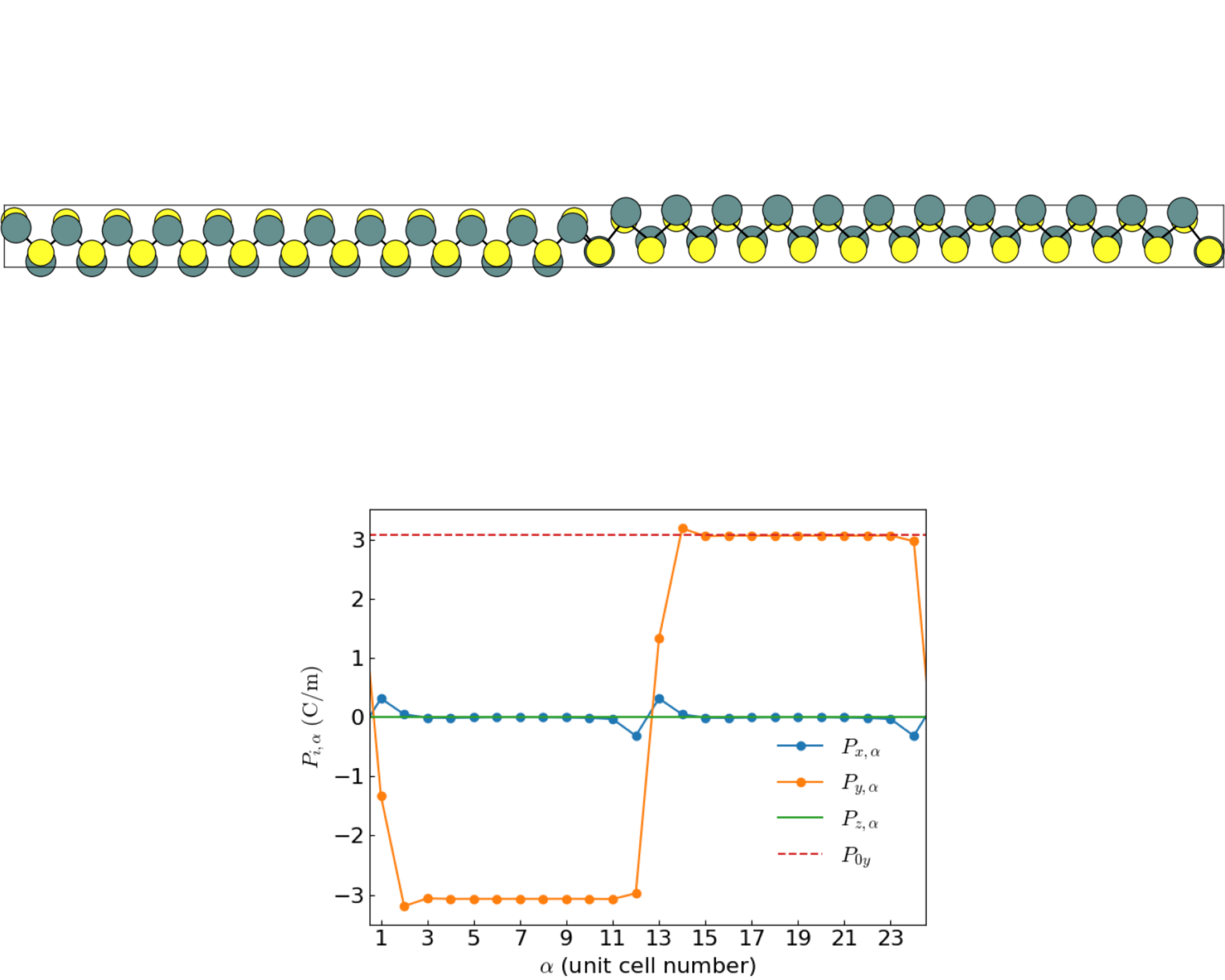}
    \centering
    \caption{
        \label{fig:atomcen}
        Top: relaxed $24\times 1\times 1$ supercell of GeS containing 2 180 degree DWs in a atom-centered configuration. It corresponds to the minimum energy 180 DW configuration of SnS and SnSe, and to the maximum in the NEB energy path of GeS and GeSe. Bottom: local polarization profile of the structure. The $x$-direction is perpendicular to the domain wall and the $y$-direction is parallel to the domain wall. }
\end{figure*}

In the case of SnS and SnSe the most stable 180 DW is the atom-centered configuration, whereas GeS and GeSe, exhibit a bond-centered configuration as the most stable DW. In figure \ref{fig:atomcen} we show an example of such a relaxed atom-centered configuration for GeS, along with its local polarization profile. A comparison with Fig. 1 in the main text illustrates the difference between the two configurations. In the calculation, the ground state of GeS and GeSe has space group symmetry P2$_1$/c (no. 14) and the ground state of SnS and SnSe (shown in figure \ref{fig:atomcen}) has space group symmetry Pma2 (no. 28).

\begin{table*}[htp]
\caption{\label{tab:tableSI}Lattice anisotropy $\delta$, spontaneous polarization calculated with the average BEC $P_0^{2D,*}$ and reference spontaneous polarization from the literature $P_0^{2D}$. We also show the diagonal components of the calculated BEC in the centrosymmetric phase $Z^{*M/X}_{\perp/\parallel}$ for each atom where $M=\text{Ge,Sn}$,  $X=\text{S,Se}$ and $\perp/\parallel$ denote directions parallel and perpendicular to the polarization.}
\begin{tabular}{lllll}
\hline
\makecell{ }&\makecell{SnSe} & \makecell{SnS}& \makecell{
GeSe}&  GeS \\
\hline
$\delta$    & 0.029 &  0.064  & 0.076 &  0.23  \\
$P_0^{2D,*}$ (pC/m) & 212 &  282   &  345  & 496   \\
$P_0^{2D}$ (pC/m)\cite{Wang_2017} & 181 &  260   &  357  & 484   \\
$Z_{\parallel}^{*M}$    & 5.71 & 5.06 &  6.82 & 6.09\\
$Z_{\perp}^{*M}$    & 5.21 & 4.30 &  5.40 &   3.58 \\
$Z_{\parallel}^{*X}$    & -5.71 & -5.06  & -6.82  & -6.09   \\
$Z_{\perp}^{*X}$    & -5.21 & -4.30 & -5.40  & -3.58 \\
%$Z_{\mathrm{M},\parallel}^{*a}$    & 5.71 & 5.06 &  6.82 & 6.09\\
%$Z_{\mathrm{M},\perp}^{*a}$    & 5.21 & 4.30 &  5.40 &   3.58 \\
%$Z_{\mathrm{X},\parallel}^{*a}$    & -5.71 & -5.06  & -6.82  & -6.09   \\
%$Z_{\mathrm{X},\perp}^{*a}$    & -5.21 & -4.30 & -5.40  & -3.58 \\
\hline
\end{tabular}
\end{table*}

\begin{table}
\caption{\label{tab:table4}Values of selected parameters in different supercells of GeS containing two 180 degree DWs. $A$ is the ratio between the polarization of the DW system far from the wall and the actual pristine monolayer polarization. $E_{form}$ is the DW formation energy per unit length. $E_{mig}$ is the DW migration energy barrier per unit length. "Full" denotes full cell relaxation, "Axis" denotes relaxation with a fixed axis parallel to the wall and "Fixed" denotes relaxation with fixed unit cell. $\eta_1$ is the strain in the long supercell direction (perpendicular to the wall) and $\eta_2$ is the strain in the short supercell direction (parallel to the wall).}
\begin{tabular}{lllllll}
\hline
\makecell{Unit\\cells}&\makecell{Cell relaxation \\constraint} & \makecell{$E_{form}$ \\(meV/Å)}& \makecell{
$E_{mig}$ \\(meV/\AA)}&  $A$    &   $\eta_1$  &  $\eta_2$  \\
\hline
& Full    & 85 &  0.035  &   0.56  &   0.06  &  -0.11 \\
 8& Axis  & 130 &  24   &  1.00  & 0.02  &  0  \\
& Fixed    &140 &  20 & 1.00  &    0 &    0 \\
\hline
& Full    & 110 & 1.0  &  0.72   &  0.04   &  -0.08    \\
 16& Axis  &139  &  24   & 1.00   & 0.01  &   0 \\
& Fixed   &140 & 22  & 1.00  &    0 &    0 \\
\hline
& Full   & 130 &  3.3  &  0.79  & 0.03  &  -0.06  \\
 24& Axis  & 150& 24  &  1.00  &  0.01 &   0 \\
& Fixed   & 150 &  23 & 1.00 &  0 &  0 \\
\hline
\end{tabular}
\end{table}

\section{Electronic properties}

\begin{figure*}
    \includegraphics[width=0.99\linewidth]{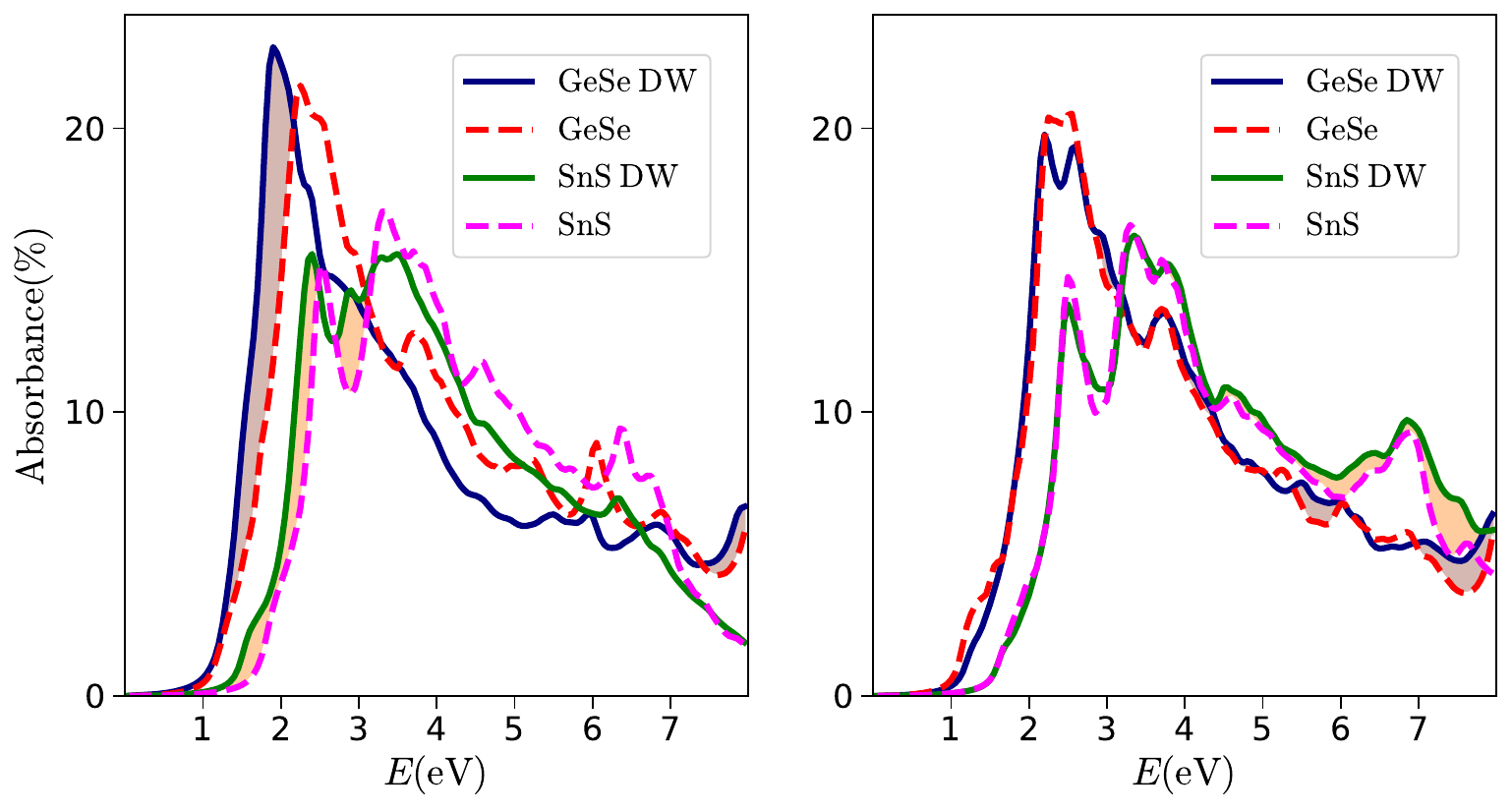}
    \centering
    \caption{
        \label{fig:absor}
        Absorbance profiles of pristine and DW configurations for GeSe and SnS. Left: $8\times1\times1$ 180 DW supercell. Right: $6\times1\times1$ 90 DW supercell.}
\end{figure*}

The absorption spectra of light polarized perpendicular to the 180 and 90 degree DW of GeSe and SnS is shown in figure \ref{fig:absor}. The structures used for the calculations contain two 180 DWs in a supercell composed of eight unit cells and two 90 DWs contained in a supercell composed of six rhombus unit cells.

\section{Coercive fields}
We consider an extended system with a single domain wall in the presence of a constant electric field $\bm{\mathcal{E}}$. In general we may write the electric enthalpy per unit cell parallel to the domain wall as
\begin{equation}
F(\{u_i^a\},\bm{\mathcal{E}}) = E(\{u_i^a\}) -\frac{1}{a_\parallel}\sum_i\mathcal{E}_iP_i(\{u_i^a\}),
\label{eq:total_energySI}
\end{equation}
where 
\begin{equation}
P_i=e\sum_{a,j}Z_{ij}^*u_j^a
\end{equation}
is the total dipole moment of the system and $a_\parallel$ is the unit cell length in the direction parallel to the domain wall. $u_i^a$ denotes the position of atom $a$, $e$ is the electronic charge and $Z_{ij}^a$ is the Born effective charge tensor of atom $a$. $E[\{u_i^a\}]$ is the electronic ground state energy per unit length with the atoms fixed at positions $u_i^a$.

We now consider the electric enthalpy along a minimum energy (NEB) path connecting an initial and final state, which differs by the domain wall having moved by a distance $d$. The path is parameterized by $\lambda$ where $\lambda=0$ corresponds to the initial state and $\lambda=1$ corresponds to the final state. Each value of $\lambda$ is characterized by a set of atomic positions $u_i^{a,\lambda}$ and thus a dipole moment $P_i(\lambda)$. The electric enthalpy can then be written as
\begin{align}
F(\lambda,\bm{\mathcal{E}}) &= E(\lambda) -\frac{1}{a_\parallel}\sum_i\mathcal{E}_iP_i(\lambda)\notag\\
&=E(\lambda)-\frac{1}{a_\parallel}\mathcal{E}P_\mathcal{E}(\lambda)
\end{align}
where $\mathcal{E}=|\bm{\mathcal{E}}|$ and $P_\mathcal{E}(\lambda)$ is the component of dipole moment parallel to the electric field. It is now convenient to choose
\begin{align}
\lambda=\frac{P_\mathcal{E}(\lambda)-P_\mathcal{E}(\lambda=0)}{P_\mathcal{E}(\lambda=1)-P_\mathcal{E}(\lambda=0)}\equiv\frac{P_\mathcal{E}(\lambda)-P_\mathcal{E}(\lambda=0)}{\Delta P_\mathcal{E}},
\label{eq:lambda}
\end{align}
where $\Delta P_\mathcal{E}$ is the change in dipole moment between the initial and final configurations on the path. The electric enthalpy may then be written as
\begin{align}
F(\lambda,\bm{\mathcal{E}})=E(\lambda)- \frac{1}{a_\parallel}\mathcal{E}\Delta P_\mathcal{E}\lambda-\frac{1}{a_\parallel}\mathcal{E}P_\mathcal{E}(\lambda=0)
\label{eq:enthalpy}
\end{align}
The system will be at equilibrium when the electric enthalpy satisfies:
\begin{equation}
\dv{ F[P_\mathcal{E},{\mathcal{E}}]}{P_\mathcal{E}}=0
\label{eq:equil0}
\end{equation}

Since $P_\mathcal{E}$ is proportional to $\lambda$, eq. \ref{eq:equil0} can be combined with eq. \ref{eq:enthalpy}. This gives the electric field required for the system to be in equilibrium at a particular point:
\begin{equation}
\mathcal{E}^\lambda = \frac{a_\parallel E'(\lambda)}{\Delta P_\mathcal{E}},
\label{eq:field}
\end{equation}
where $E_0'$ denotes derivative with respect to $\lambda$.
The coercive field corresponds to the force provided by the electric field being larger than the maximum slope of $E(\lambda)$ and thus 
\begin{equation}
\mathcal{E}_\mathrm{C}=a_\parallel\frac{\mathrm{max}\big[ E'(\lambda)\big] }{\Delta P_\mathcal{E}}
\end{equation}

We now consider an applied field at an angle $\theta$ with respect to the domain wall. For the 180 domain wall the change in dipole moment (along the field) between initial and final states is then given by $\Delta P_\mathcal{E}=2ab\cos{\theta}P_0^\mathrm{2D}$, where $P_0^\mathrm{2D}$ is the spontaneous 2D polarization. Since in this case $a_\parallel=a$ the coercive field becomes 
\begin{equation}
\mathcal{E}_\mathrm{C}^{180}(\theta)=\frac{\mathrm{max}\big[ E'(\lambda)\big] }{2b\cos{\theta}P_0^\mathrm{2D}}.
\end{equation}
In the case of a 90 degree domain wall, the actual DW angle deviates slightly from 90 degrees as shown in the main text. We denote this DW angle as $\gamma$. We have $\Delta P_\mathcal{E}=4ab\cos(\theta-\gamma/2)P_0^\mathrm{2D}$ and $a_\parallel=\sqrt{a^2+b^2}$. We thus get
\begin{align}
\mathcal{E}_\mathrm{C}^{90}(\theta)&=\frac{\sqrt{a^2+b^2}\mathrm{max}\big[E'(\lambda)\big]}{4ab\cos(\theta-\gamma/2)P_0^\mathrm{2D}}\notag\\
&=\frac{\mathrm{max}\big[E'(\lambda)\big]}{4b\cos(\gamma/2)\cos(\theta-\gamma/2)P_0^\mathrm{2D}}
\end{align}
and it is clear that the smallest field corresponds to $\theta=\gamma/2$. That is, the field should be parallel to the polarization.

We note that in standard NEB calculations one typically obtains the minimum energy path in terms of a parameter $\tilde\lambda$ that measures the normalized deviations of atomic positions along the path. For the calculations in the present work we checked that the deviations between $\tilde\lambda$ and $\lambda$ defined in Eq. \eqref{eq:lambda} is negligible and we have simply used $\tilde\lambda$ for our calculations.

It is often convenient to resolve the total dipole moment of a big supercell in terms of contributions from individual unit cells. The total dipole moment is then written as
\begin{equation}
P_i=\sum_\alpha P_i^\alpha
\end{equation}
where
\begin{equation}
P_i^\alpha=e\sum_{a\in\alpha,j}Z_{ij}^*u_j^a
\end{equation}
is the dipole contribution from unit cell $\alpha$. For macroscopic modelling the polarization may be regarded as a continuous field and the electric enthalpy in Eq. \eqref{eq:total_energySI} is written as
\begin{align}
F(\{u_i^a\},\bm{\mathcal{E}}) &= E(\{u_i^a\}) -\frac{1}{a_\parallel}\sum_{i}\mathcal{E}_i\sum_{\alpha}P_i^\alpha\notag\\
&= E(\{u_i^a\}) -\frac{1}{a_\parallel a_\perp}\sum_{i}\mathcal{E}_i\sum_{\alpha}a_\perp P_i^\alpha\notag\\
&\approx E[P^\mathrm{2D}_i(x)] -\sum_{i}\mathcal{E}_i\int dx P_i^\mathrm{2D}(x),
\end{align}
where we assumed that $P_i^\alpha$ varies on much larger length scales than $a_\perp$ and defined $P^\mathrm{2D}_i(x)\approx P_i^\alpha/a_\perp a_\parallel$.

\providecommand{\noopsort}[1]{}\providecommand{\singleletter}[1]{#1}%
\providecommand{\newblock}{}

\end{document}